\documentclass[conference]{IEEEtran}
\IEEEoverridecommandlockouts
\usepackage{cite}
\usepackage{amsmath,amssymb,amsfonts}
\usepackage{algorithmic}
\usepackage{graphicx}
\usepackage{textcomp}
\usepackage{xcolor}
\usepackage{hyperref}
\hypersetup{
    colorlinks=true,
    linkcolor=blue,
    filecolor=blue,      
    urlcolor=blue,
    pdftitle={JCET2022},
    pdfpagemode=FullScreen,
}
\usepackage{multirow, multicol}
\def\BibTeX{{\rm B\kern-.05em{\sc i\kern-.025em b}\kern-.08em
    T\kern-.1667em\lower.7ex\hbox{E}\kern-.125em}}
\begin{document}

\title{Integrating Fourier Transform and Residual Learning for Arctic Sea Ice Forecasting
}

\author{\IEEEauthorblockN{Louis Lapp}
\IEEEauthorblockA{\textit{Baltimore Polytechnic Institute}\\
Baltimore, MD, USA \\
2louislapp@gmail.com}
\and
\IEEEauthorblockN{Sahara Ali}
\IEEEauthorblockA{
\textit{University of Maryland Baltimore County
}\\
Baltimore, MD, USA \\
sali9@umbc.edu}
\and
\IEEEauthorblockN{Jianwu Wang}
\IEEEauthorblockA{
\textit{University of Maryland Baltimore County}\\
Baltimore, MD, USA \\
jianwu@umbc.edu}
}

\maketitle

\begin{abstract}
Arctic sea ice plays integral roles in both polar and global environmental systems, notably ecosystems, communities, and economies. As sea ice continues to decline due to climate change, it has become imperative to accurately predict the future of sea ice extent (SIE). Using datasets of Arctic meteorological and SIE variables spanning 1979 to 2021, we propose architectures capable of processing multivariate time series and spatiotemporal data. Our proposed framework consists of ensembled stacked Fourier Transform signals (FFTstack) and Gradient Boosting models. In FFTstack, grid search iteratively detects the optimal combination of representative FFT signals, a process that improves upon current FFT implementations and deseasonalizers. An optimized Gradient Boosting Regressor is then trained on the residual of the FFTstack output. Through experiment, we found that the models trained on both multivariate and spatiotemporal time series data performed either similar to or better than models in existing research. In addition, we found that integration of FFTstack improves the performance of current multivariate time series deep learning models. We conclude that the high flexibility and performance of this methodology have promising applications in guiding future adaptation, resilience, and mitigation efforts in response to Arctic sea ice retreat.
\end{abstract}

\begin{IEEEkeywords}
Arctic sea ice extent, Fourier Transform, machine learning, time series forecasting
\end{IEEEkeywords}

\section{Introduction}

\begin{table*}[htbp]
\caption{\label{table:datasets-overview}Overview of Datasets}
\begin{center}
\begin{tabular}{|p{0.125\linewidth}|p{0.125\linewidth}|p{0.01\linewidth}|p{0.065\linewidth}|p{0.115\linewidth}|p{0.275\linewidth}|}
\hline
\multicolumn{1}{|c|}{\textbf{Dataset}}& \multicolumn{1}{c|}{\textbf{Domain}}& \multicolumn{1}{c|}{\textbf{Interval}}& \multicolumn{1}{c|}{\textbf{Shape}}& \multicolumn{1}{c|}{\textbf{Shape details}}& \multicolumn{1}{c|}{\textbf{Description}} \\
\hline
2D time series \cite{ali2021sea} & 01/1979 to 08/2021 & Monthly & (512, 10) & (time, variable) & Multivariate time series of 9 Arctic meteorological variables and sea ice extent \\
\hline
3D time series \cite{seaiceindexv3} & 01/1979 to 08/2021 & Monthly & (512, 448, 304) & (time, longitude, latitude) & Spatiotemporal time series of SIE images \\
\hline
\end{tabular}
\end{center}
\end{table*}

\begin{table}[htbp]
\caption{\label{table:dataset-variables}2D Dataset Variables}
\begin{center}
\begin{tabular}{|c|c|c|}
\hline
\multicolumn{1}{|c|}{\textbf{Variable}}& \multicolumn{1}{c|}{\textbf{Range}}& \multicolumn{1}{c|}{\textbf{Unit}} \\
\hline
Air temperature & [200, 350] & K \\
\hline
Sea surface temperature & [200, 350] & K \\
\hline
Surface pressure & [400, 1100] & hPa \\
\hline
Specific humidity & [0, 0.1] & kg/kg \\
\hline
Shortwave down radiation & [0, 1500] & W/m\textsuperscript{2} \\
\hline
Longwave down radiation & [0, 700] & W/m\textsuperscript{2} \\
\hline
Rain rate & [0, 800] & mm/day \\
\hline
Snowfall rate & [0, 200] & mm/day \\
\hline
Wind velocity & [0, 40] & m/s \\
\hline
Sea ice extent & [4, 13] & million km\textsuperscript{2} \\
\hline
\end{tabular}
\end{center}
\end{table}
Arctic sea ice is a critical component of polar atmospheric and oceanic systems. Yet, our linear regression analysis reveals marked sea ice decline in the past several decades, averaging 0.081 million km\textsuperscript{2} of loss per year. As a matter of fact, the Arctic has undergone warming at globally unprecedented rates, a phenomenon known as Arctic amplification \cite{vihma2014effects}. Once Arctic temperature reaches the bifurcation point, estimated around 2°C, summer sea ice may be irreversibly lost \cite{mahlstein2012september}. However, the timeline of this decline is uncertain due to the chronic underestimation of current predictions \cite{stroeve2007arctic}. The effects on the Arctic environment are numerous, including increases in local sea level, air temperatures, and moisture levels \cite{casas2020sea, light2008transmission}. Causal sequences have also been linked between sea ice loss, anomalous atmospheric patterns, and extreme weather events in Europe, Asia, and North America \cite{wu2013relationship, hu2010tundra}.

Arctic ecosystems are directly impacted by the geophysical effects of sea ice loss, leading to plankton growth, and polar bear vulnerability \cite{zhang2010modeling, regehR2010survival}. Humans have also been impacted, putting at risk the livelihood of indigenous peoples, and increasing shipping route accessibility \cite{pearce2010inuit, stephenson2013projected}. Thus, Arctic sea ice loss is a prominent issue for people and ecosystems on both the polar and global levels, thereby warranting further research.

\subsection{Related Work}
The majority of current SIE prediction models rely on statistical, dynamical, machine learning (ML), and more recently, deep learning (DL) techniques. For reference, we use one month lead time Root Mean Square Error (RMSE) score as the evaluation metric for comparison. Using daily sea ice concentration data, Wang et al. implemented a vector autoregression technique to predict sea ice concentration and derived September mean SIE, achieving an RMSE of 0.45 million km\textsuperscript{2} \cite{wang2016predicting}. Using atmospheric and oceanic variables, Wang et al. implemented the Climate Forecast System (CFSv2), a fully-coupled physical model, to predict Arctic SIE, achieving an RMSE range of 0.2-0.6 million km\textsuperscript{2} \cite{wang2013seasonal}. Using atmospheric and oceanic variables, Ali et al. implemented an ensemble of Long-Short Term Memory (LSTM) models to predict Arctic SIE, achieving an RMSE of 0.586 million km\textsuperscript{2} \cite{ali2021sea}. Using a similar dataset, Ali et al. also conducted comparative analyses of ML and probabilistic models, finding that a Multiple Linear Regression model performed best with an RMSE of 0.433 million km\textsuperscript{2} for predicting one month ahead SIE \cite{ali2022benchmarking}. Kim et al. proposed a multitask Convoluted Neural Network (CNN) LSTM technique using atmospheric, oceanic, and pixel-wise sea ice concentrations. Their DL model predicts derived Arctic SIE from concentration, achieving an RMSE of 0.303 million km\textsuperscript{2} \cite{kim2021multi}. Overall, DL has demonstrated significant potential in SIE prediction. However, the enlisted models fail to directly take advantage of intra-yearly and inter-yearly trends, likely contributing to poor extremum accuracy and inadequate intra-extrema performance. In addition, only select models have SIE visualization capabilities, and none have implemented an algorithm to remove cyclic trends.

In light of these shortcomings, this study applies climatic trends and ML to predict Arctic SIE one month in advance with comparable or improved results to current models. 
This paper has the following contributions: 1) We propose FFTstack, an iterative Fourier Transform-based system for modeling cyclic temporal trends; 2) We further propose streamlined architectures, which accept both multivariate time series and spatiotemporal data, for ensembling FFTstack with linear regression, ML, and/or existing research models; 3) We perform a detailed comparative analysis of our proposed models and those produced by existing research. Our implementation code can be accessed at the GitHub repository\footnote{\href{https://github.com/big-data-lab-umbc/sea-ice-prediction/tree/main/fftstack-icmla}{github.com/big-data-lab-umbc/sea-ice-prediction/tree/main/fftstack-icmla}}.

\section{Datasets and Problem Statement}
\subsection{Datasets}
To forecast SIE, we use two datasets of varying shapes and structures. The multivariate architecture uses the same multivariate time series as in \cite{ali2021sea} that spans 01/1979 to 08/2021. It includes daily entries (n=15584) of nine Arctic atmospheric and oceanic variables, and sea ice extent, as enlisted in Tables \ref{table:datasets-overview} and \ref{table:dataset-variables}. Here, the target variable is sea ice extent. The monthly (n=512) and yearly (n=43) multivariate time series were derived from the daily data through averaging. For brevity, ‘2D data’ refers to the monthly version of this multivariate time series, and '2D architecture' refers to the architecture capable of processing 2D data.

The 3D architecture used the Monthly Sea Ice Index, a spatiotemporal time series, provided by the National Snow and Ice Data Center in \cite{seaiceindexv3}. For this application, SIE images were used, with a concentration threshold of 15\% and spatial resolution of 25km x 25km. The images are of shape (448, 304). This dataset spans the same time period as the previous. For brevity, ‘3D data’ refers to this spatiotemporal time series, and '3D architecture' refers to the architecture capable of processing 3D data.

\subsection{Problem Statement}
Given \textit{N} months of historic meteorological and sea-ice data \textit{X}, learn a function to forecast sea ice extent \textit{Y} for the next one month, as shown in Equation \eqref{eq1} where $t$ $\epsilon$ $(1,2,...,N)$.
\begin{equation}
    Y_{t+1}=f(X_{t-n}, X_{t-n+1},…, X_{t})\label{eq1}
\end{equation}

\section{Methods}
\begin{figure}[htbp]
\centerline{\includegraphics[width=0.7\linewidth]{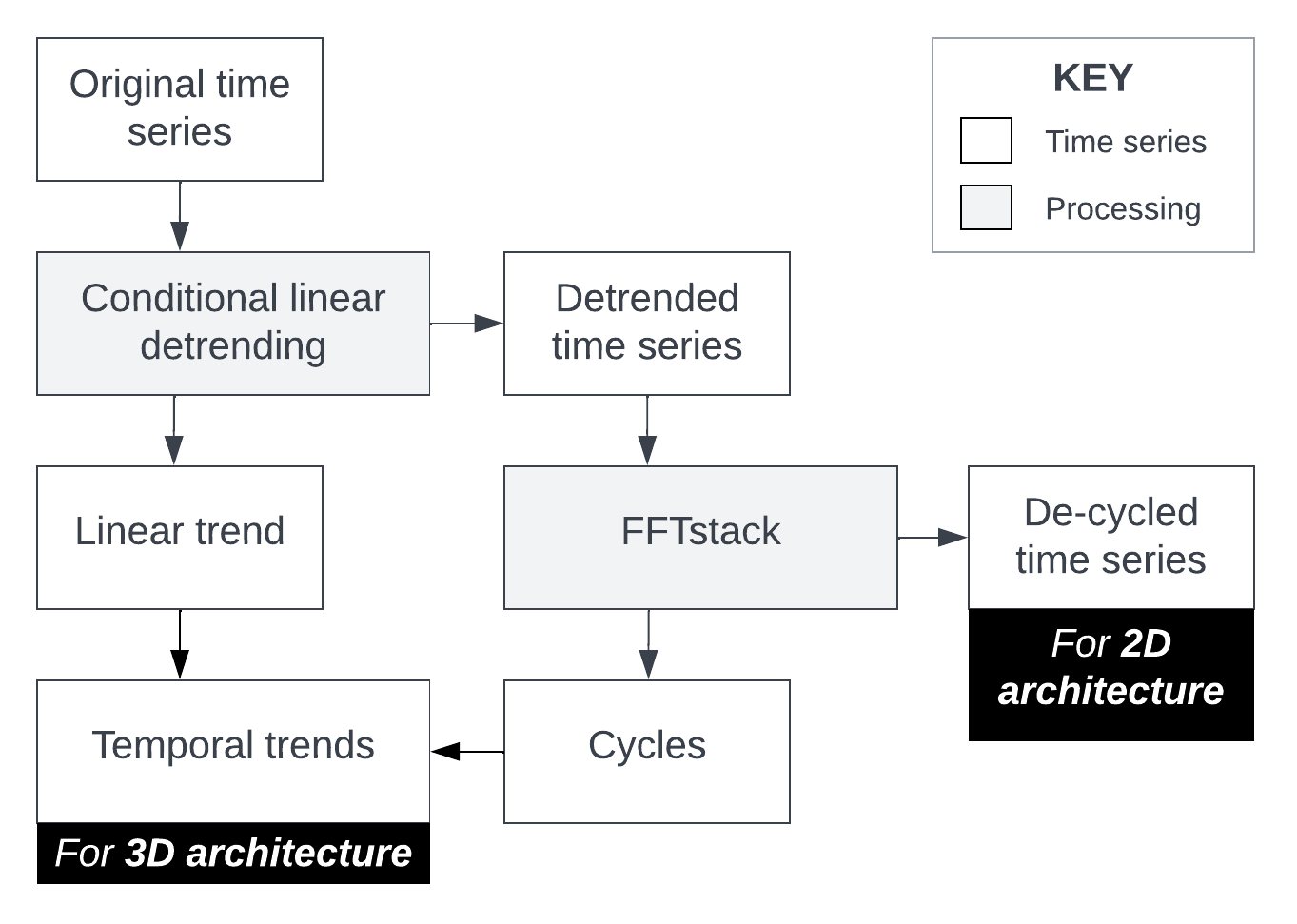}}
\caption{Overview of Methodology}
\label{fig:methods-overview}
\end{figure}

This study evaluates the potential of leveraging temporal trends in data to predict Arctic SIE numerically and visually. First, we propose FFTstack, a novel approach to using Fourier Transform for modeling cyclic trends. Then, we explore the potential for applying FFTstack to Arctic sea ice prediction via integration with streamlined architectures. For the 2D architecture, we evaluate the potential to train an ML model on the deviations from the FFTstack-detected trends. We generate various baseline 2D models for comparison. For the 3D architecture, we apply FFTstack in a pixel-wise manner. A general overview of these processes is shown in Figure \ref{fig:methods-overview}.

\subsection{Data Preprocessing}
To begin, both 2D and 3D data were split into 80:20 ratio, where approximately 80\% was reserved for training (n=408), i.e. from 01/1979 to 12/2012, and 20\% was set aside for testing (n=104), i.e. from 01/2013 to 08/2021.

Each monthly time series variable, \textit{x}, underwent conditional detrending, collectively forming 'detrended data'. First, a linear regression model was trained on the yearly time series, \textit{x}\textsubscript{year}. If the trend was statistically significant (\(p<0.05\)), the linear regression model was trained on the monthly time series. The trend was subtracted from the monthly time series to obtain a detrended time series for that variable. If the trend was not statistically significant, the average of the monthly time series was subtracted from the monthly time series itself to center it around zero. The final trend, whether linear regression or average, is denoted by \textit{l}. Thus, only variables with statistically significant inter-yearly trends were detrended, thereby preventing variables with trends occurring by chance from being detrended as well, as shown in Equation \eqref{eq2}. 
\begin{equation}
    l_x = \begin{cases}
    linreg\left(x\right) & ,p\left(linreg(x_{year})\right)<0.05 \\
x $-$ \bar{x} & ,p\left({linreg(x}_{year})\right)\ge0.05
    \end{cases}
\label{eq2}
\end{equation}

\subsection{FFTstack}
\label{fftstack}
Fast Fourier Transform (FFT) and Inverse Fast Fourier Transform (IFFT) are computationally efficient implementations of mathematical transforms that conduct analysis from time to frequency and frequency to time domains, respectively.

\subsubsection{Composite Fourier Transform Model}
Individual FFT-based cycles, $c$, were generated with an FFT-based model for the monthly detrended time series of every variable, $x$. Two parameters were specified per variable: number of iterations, $a$, and threshold, $b$. The following process was iterated $a$ times. First, FFT was conducted on the detrended time series. The resulting complex frequency domain values were converted to real numbers and normalized. By scaling FFT output to a set range between 0 and 1, normalization eased the detection of a threshold.

Next, a binary mask of booleans removed noise from the FFT output, where any frequencies with a normalized amplitude greater than or equal to the specified threshold were assigned 1, while all others were assigned 0. Then, the binary mask was applied to the FFT frequencies by element-wise multiplication. 
IFFT was subsequently conducted on the cleaned FFT frequencies, effectively returning cycles that modeled existing oscillation. If the number of iterations exceeded one (\(a>1\)), the process would repeat with the residual of the cycles serving as the new input, thereby removing leftover oscillation from the previous iteration. The sequential removal of trends in these residuals was more robust against overfitting the FFT model than determining a more precise threshold. Thus, a combination of the two was used instead, as illustrated in Equation \eqref{eq3}.
\begin{multline}
    \label{eq3}
    c_x = \sum_{i=1}^a\text{IFFT}\left(\text{FFT}\left(x-\sum_{j=1}^ic_j\right) \right.\\
    \left.\ast\left[\text{norm}\left(\left|\text{FFT}\left(x-\sum_{j=1}^ic_j\right)\right|\right)\ge b\right]\right)
\end{multline}

FFTstack returned either the sum of all cycles generated by each iteration, called 'composite cycles', or the residual of the composite cycles, called 'de-cycled data'. The 2D architecture will use the de-cycled data, while the 3D architecture will use the composite cycles instead.

\begin{table*}[htbp]
\caption{\label{table:search-spaces}Hyperparameter Search Spaces}
\begin{center}
\begin{tabular}{|c|c|c|c|}
\hline
\multicolumn{1}{|c|}{\textbf{Model}}& \multicolumn{1}{c|}{\textbf{Used by}}& \multicolumn{1}{c|}{\textbf{Parameter}}& \multicolumn{1}{c|}{\textbf{Values}} \\
{} & {} & {} & Format: (start, end, step=1) or [item\_1, ... , item\_n] \\
\hline
Composite FFT & FFTstack, 2D/3D architectures & n\_iteration & (1, 5) \\
\cline{3-4}
{} & {} & threshold & (0.1, 0.9, 0.1) \\
\hline
Gradient Boosting & 2D architecture only & n\_estimators & (25, 500, 25) \\
\cline{3-4}
{} & {} & max\_depth & (1, 5) \\
\cline{3-4}
{} & {} & learning\_rate & [1, 0.5, 0.1, 0.05, 0.01] \\
\cline{3-4}
{} & {} & lags & (12, 60, 12) \\
\hline
\end{tabular}
\end{center}
\end{table*}

\subsubsection{Grid search}
Grid search was employed to optimize the FFT-based model. First, detrended was further split into an 80:20 ratio, with approximately 80\% reserved for training (n=324), i.e. from 01/1979 to 12/2005, and 20\% reserved for validation (n=84), i.e. from 01/2006 to 12/2012. Due to the generally strong periodicity of Arctic meteorological variables and conservative modeling technique of FFTstack, additional splitting was not considered necessary to ensure a well-fit model, as substantiated by Section \ref{2dfftstackresults}. Then, the algorithm determined the parameter combination with the lowest residual metric for the training set. The trend in the training data was extrapolated to the testing data and de-cycled accordingly. The search space encompassed a number of iterations and threshold over 45 combinations, as shown in Table \ref{table:search-spaces}.

\subsubsection{Residual metric}
The residual metric, \textit{RM}, was designed to evaluate the size of the residual while preventing overfitting. Median Absolute Deviation (MAD) was selected as a robust measure of variability. The percent change between the MAD of the train and validation partition was calculated. If the percent change was greater than 0.1, the cycles were considered overfit and the metric was set to infinity. Otherwise, the metric was set equal to the average MAD of both the train and test time series, as shown in Equation \eqref{eq4}. Because lower variability implied a smaller residual, a lower residual metric was desirable.
\begin{multline}
    RM=\left[\frac{MAD\left(x_{train}\right)-MAD\left(x_{val}\right)}{MAD\left(x_{train}\right)}\le0.1\right] \\
    \ast\left[\frac{MAD\left(x_{train}\right)+MAD\left(x_{val}\right)}{2}\right]\label{eq4}
\end{multline}
\begin{table*}[htbp]
\caption{\label{table:architecture-summaries}Summary of 2D and 3D Architectures}
\begin{center}
\begin{tabular}{|c|c|c|}
\hline
\multicolumn{1}{|c|}{\textbf{Model}}& \multicolumn{1}{c|}{\textbf{Type}}& \multicolumn{1}{c|}{\textbf{Architecture}} \\
\hline
2D GBR & Baseline & 2D data → GBR → SIE numerical prediction\\
\hline
2D GBR (detrended) & Baseline & 2D data → linear detrending + GBR → SIE numerical prediction \\
\hline
2D FFTstack & Proposed & 2D data → linear detrending + FFTstack → Cyclic trends \\
\hline
2D GBR (de-cycled) & Proposed & 2D data → linear detrending + FFTstack + GBR → SIE numerical prediction \\
\hline
3D FFTstack & Proposed & 3D data → FFTstack → SIE image prediction \\
\hline
\end{tabular}
\end{center}
\end{table*}

\subsection{Architectures}
Two architectures were developed to accommodate the differing shapes and structures of SIE data. An overview of these architectures is available in Table \ref{table:architecture-summaries}.
\subsubsection{2D architecture}
\label{2darchitecture}
The 2D architecture accepts multivariate time series data for univariate time series prediction of SIE. To accomplish this, an ML model was trained on the FFTstack-generated residuals, a process referred to as ‘residual learning’ (not to be confused with the DL term). We used Gradient Boosting Regressors (GBR), a type of ML algorithm that ensembles weak learners, due to their capacity for predicting complex patterns such as those found in the pattern-neutral residuals.

The proposed 2D GBR (de-cycled) model accepted 2D de-cycled data explained in Section \ref{fftstack} as input. Two baseline models were also generated for comparison, with the 2D GBR (original) model and 2D GBR (detrended) model accepting unaltered 2D data and 2D detrended data, respectively.

First, an expanding window splitter with two folds was applied to the inputted training data to ensure consistent performance. In Fold 1, the 2D data was split into a 2:3 ratio, where approximately 66\% was reserved for training (n=204), i.e. from 01/1979 to 12/1995, and 33\% was set aside for validation (n=104), i.e. from 01/1996 to 08/2004. In Fold 2, the 2D data was split into a 3:4 ratio, where approximately 75\% was reserved for training (n=308), i.e. from 01/1979 to 08/2004, and 25\% was set aside for validation (n=100), i.e. from 09/2004 to 12/2012.

Then, grid search optimized a recursive autoregressive GBR for lowest Root Mean Squared Error (RMSE), determining optimal hyperparameter and lag combinations. The search spaces are provided in Table \ref{table:search-spaces}.

As denoted in Equation \eqref{eq5}, the final prediction of the 2D GBR (de-cycled) model was obtained by adding the linear trend, $L$, composite cycles, $C$, and optimized GBR predictions based on $n$ past months of de-cycled data, $R$. 
\begin{equation}
    Y_{t+1}=(L+C)_{t+1}+GBR(R_{t-n}, R_{t-n+1}, R_{t})\label{eq5}
\end{equation}

\subsubsection{3D architecture}
The 3D architecture applies FFTstack to images to forecast the temporal trends contained in each pixel. First, the 3D data was flattened and transposed, such that the training and testing data had shapes of (136192, 408) and (136192, 104), respectively. Then, FFTstack was applied to each of the 136192 pixel time series, with cycles of temporal trends returned as the final output. Finally, all pixel time series predictions were rearranged into a 3D shape, such that the outputted training and testing forecasts occurred in shapes of (408, 448, 304) and (104, 448, 304), respectively.

\subsection{FFTstack Integration with Current Research}
To evaluate its applicability to existing research, FFTstack was integrated into the frameworks of \cite{ali2021sea} and \cite{ali2022benchmarking}. For both, FFTstack was applied to the input data. Then, the existing models were trained on the de-cycled data in accordance with the residual learning framework entailed in Section \ref{2darchitecture}.

\section{Results}
We present results from two architectures encompassing five models which ensemble various linear regression, FFTstack, and ML algorithms. In addition, we evaluate the effect of integrating FFTstack with models from existing research.
\begin{table*}[htbp]
\caption{\label{table:transformer-results}Detrender and FFTstack Model Statistics and Parameter Configurations}
\begin{center}
\begin{tabular}{|c|c|c|c|c|c|}
\hline
\multicolumn{1}{|c|}{\textbf{Variable}}& \multicolumn{1}{c|}{\textbf{Detrender}}& \multicolumn{4}{|c|}{\textbf{FFTstack}} \\
\cline{2-6}
\multicolumn{1}{|c|}{}& \multicolumn{1}{c|}{\textbf{\textit{P-value}}}& \multicolumn{1}{c|}{\textbf{\textit{Parameters}}}& \multicolumn{1}{c|}{\textbf{\textit{\% change}}}& \multicolumn{1}{c|}{\textbf{\textit{Residual metric}}}& \multicolumn{1}{c|}{\textbf{\textit{\% of Initial}}} \\
\hline
Air temperature & $<$0.001 & \{‘n\_iteration’: 2, ‘threshold’: 0.10\} & 0.079 & 0.594 & 0.054891 \\
\hline
Sea surface temperature & $<$0.001 & \{‘n\_iteration’: 1, ‘threshold’: 0.05\} & 0.077 & 0.159 & 0.201613 \\
\hline
Surface pressure & 0.75 & \{‘n\_iteration’: 1, ‘threshold’: 0.25\} & 0.049 & 1.574 & 0.786065 \\
\hline
Specific humidity & $<$0.001 & \{‘n\_iteration’: 2, ‘threshold’: 0.10\} & 0.078 & 0.069 & 0.083597 \\
\hline
Shortwave down radiation & 0.25 & \{‘n\_iteration’: 2, ‘threshold’: 0.05\} & -0.044 & 0.685 & 0.009984 \\
\hline
Longwave down radiation & $<$0.001 & \{‘n\_iteration’: 4, ‘threshold’: 0.55\} & 0.009 & 1.903 & 0.043795 \\
\hline
Rain rate & 0.03 & \{‘n\_iteration’: 2, ‘threshold’: 0.10\} & 0.058 & 0.068 & 0.244662 \\
\hline
Snowfall rate & 0.007 & \{‘n\_iteration’: 4, ‘threshold’: 0.85\} & 0.094 & 0.047 & 0.257701 \\
\hline
Wind velocity & 0.38 & \{‘n\_iteration’: 4, ‘threshold’: 0.75\} & -0.030 & 0.083 & 0.269971 \\
\hline
Sea ice extent & $<$0.001 & \{‘n\_iteration’: 2, ‘threshold’: 0.25\} & 0.081 & 284681.194 & 0.099539 \\
\hline
\end{tabular}
\end{center}
\end{table*}

\subsubsection{Performance metrics}
To evaluate the SIE prediction capability of both study and out-of-study models, performance metrics including RMSE, Normalized Root Mean Squared Error (NRMSE), calculated by dividing RMSE by the mean of the dataset, and R\textsuperscript{2} were used. To evaluate long-term relevancy, we calculated the p-value of a linear regression model trained on the averaged yearly prediction RMSE. A non-statistically significant trend (\(p\ge0.05\)) indicated the model maintained consistent performance over time.

The MAD percentage of initial was specifically used for the FFTstack model to evaluate the size of the residual relative to the input, as shown in Equation \eqref{eq6}.
\begin{equation}
    \% \ of \ initial=\frac{MAD(x-[l_x+c_x])}{MAD(x)} \label{eq6}
\end{equation}

A smaller percentage suggested FFTstack removed more oscillation, implying a superior fit. Thus, FFTstack removed most oscillation from variables with a percentage less than 0.1, whereas only adequate removal occurred for those with a percentage less than 0.5. A percentage outlying above 0.75 suggested that FFTstack removed minor oscillation.


\subsubsection{2D FFTstack}
\label{2dfftstackresults}
The variables enlisted in Table \ref{table:transformer-results} exhibited statistically meaningful linear trends and were detrended. All variables had at least one parameter combination that attained a MAD percent change between training and validation less than 0.10, suggesting the optimal FFTstack models were not overfit. FFTstack demonstrated particularly strong performance for five variables, but poor performance for surface pressure, likely due to its low periodicity.


\begin{table*}[htbp]
\caption{\label{table:ml-results}2D Gradient Boosting Regressor Parameter Configurations and Performance Statistics}
\begin{center}
\begin{tabular}{|c|c|c|c|c|c|}
\hline
\multicolumn{1}{|c|}{\textbf{Model}}& \multicolumn{3}{|c|}{\textbf{Train}}& \multicolumn{2}{|c|}{\textbf{Test}} \\
\cline{2-6}
\multicolumn{1}{|c|}{}& \multicolumn{1}{c|}{\textbf{\textit{Hyperparameters}}}& \multicolumn{1}{c|}{\textbf{\textit{Lags}}}& \multicolumn{1}{c|}{\textbf{\textit{RMSE}}}& \multicolumn{1}{c|}{\textbf{\textit{RMSE}}}& \multicolumn{1}{c|}{\textbf{\textit{R\textsuperscript{2}}}} \\
\hline
GBR (original) & \{‘learning\_rate’: 0.1, 'max\_depth': 2, 'n\_estimators': 325\} & 36 & 565069.333 & 637720.768 & 0.965 \\
\hline
GBR (detrended) & \{'learning\_rate': 0.05, 'max\_depth': 2, 'n\_estimators': 275\} & 12 & 409503.288 & 475498.916 & \textbf{0.981} \\
\hline
GBR (de-cycled) & \{'learning\_rate': 0.05, 'max\_depth': 3, 'n\_estimators': 175\} & 12 & 358979.165 & \textbf{298025.108} & 0.610 \\
\hline
\end{tabular}
\end{center}
\end{table*}

\subsubsection{Architectures}
As shown in Table \ref{table:ml-results}, each optimized 2D model contained a lag value consistent with a year, suggesting the hyperparameter search space was sufficiently large. The baseline and test model were both slightly overfit, exhibiting higher test RMSE than train RMSE. The de-cycled model was well-fit, exhibiting a test RMSE less than train RMSE. The detrended and baseline model achieved successively higher R\textsuperscript{2}. The de-cycled model achieved the lowest R\textsuperscript{2}, likely due to the noisy nature of its input.

\begin{table*}[htbp]
\caption{\label{table:performance-comparison}Comparative Performance of Study and Existing Research Models}
\begin{center}
\begin{tabular}{|c|c|c|c|c|c|c|}
\hline
\multicolumn{1}{|c|}{\textbf{Model}}& \multicolumn{1}{c|}{\textbf{Description}}& \multicolumn{1}{c|}{\textbf{Data}}& \multicolumn{4}{|c|}{\textbf{Testing Performance}} \\
\cline{4-7}
\multicolumn{1}{|c|}{}& \multicolumn{1}{|c|}{}& \multicolumn{1}{|c|}{}& \multicolumn{1}{c|}{\textbf{\textit{RMSE (mil. km\textsuperscript{2})}}}& \multicolumn{1}{c|}{\textbf{\textit{NRMSE}}}& \multicolumn{1}{c|}{\textbf{\textit{R\textsuperscript{2}}}}& \multicolumn{1}{c|}{\textbf{\textit{P-value}}} \\
\hline
Wang et al. \cite{wang2016predicting} & Statistical model & Meteorological variables & 0.45 & - & - & - \\
\hline
Wang et al. \cite{wang2013seasonal} & Physical model & Meteorological variables & 0.2–0.6 & - & - & - \\
\hline
Ali et al. \cite{ali2021sea} & LSTM model & 10 meteor. and SIE vars. & 0.660 & 0.0628 & 0.963 & 0.026 \\
\hline
\cite{ali2021sea} with FFTstack & LSTM + FFTstack & 10 meteor. and SIE vars. & 0.412 & 0.0400 & 0.986 & 0.64 \\
\hline
Ali et al. \cite{ali2022benchmarking} & Linear regression & 11 meteor. and SIE vars. & 0.453 & 0.0431 & 0.982 & 0.063 \\
\hline
\cite{ali2022benchmarking} with FFTstack & Lin. reg. + FFTstack & 11 meteor. and SIE vars. & 0.356 & 0.0346 & 0.990 & 0.72 \\
\hline
Kim et al. \cite{kim2021multi} & CNN/LSTM model & 10 meteor. and SIE vars. and SIE images & 0.303 & 0.0292 & - & - \\
\hline
3D FFTstack & See Table \ref{table:architecture-summaries} & SIE images & 0.361 & - & - & 0.075 \\
\hline
2D GBR (original) & See Table \ref{table:architecture-summaries} & 9 meteor. vars. and SIE & 0.638 & 0.0604 & 0.965 & 0.015 \\
\hline
2D GBR (detrended) & See Table \ref{table:architecture-summaries} & 9 meteor. vars. and SIE & 0.475 & 0.0451 & 0.981 & 0.49 \\
\hline
2D FFTstack & See Table \ref{table:architecture-summaries} & SIE variable & 0.502 & 0.0475 & 0.978 & 0.14 \\
\hline
2D GBR (de-cycled) & See Table \ref{table:architecture-summaries} & 9 meteor. vars. and SIE & \textbf{0.298} & \textbf{0.0282} & \textbf{0.992} & 0.29 \\
\hline
\end{tabular}
\end{center}
\end{table*}

\begin{figure}[htbp]
\centerline{\includegraphics[width=\linewidth]{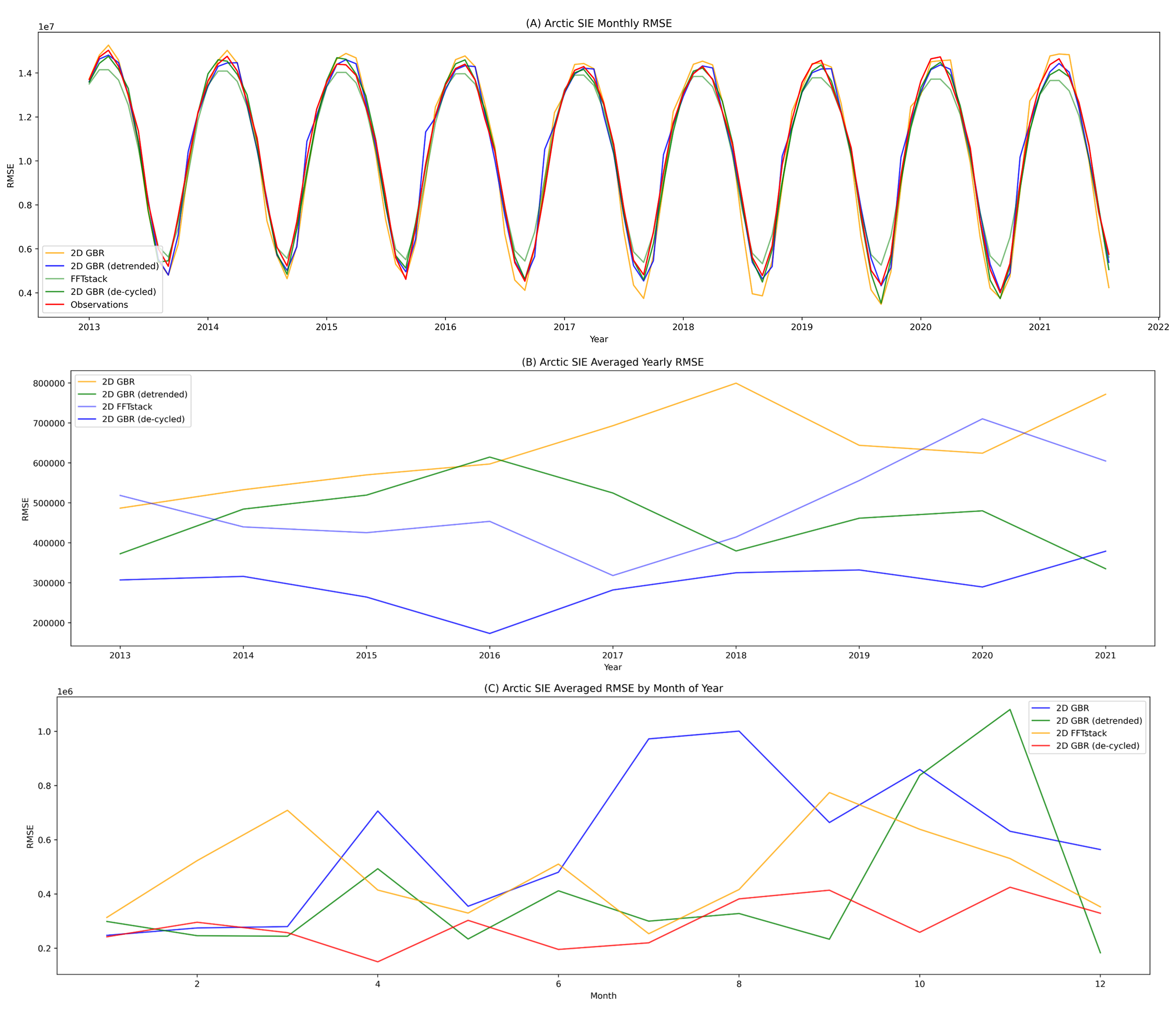}}
\caption{Comparative RMSE analysis by (A) month, (B) year, and (C) month of year for 2D models}
\label{fig:SIEgraphs}
\end{figure}

\begin{figure}[htbp]
\centerline{\includegraphics[width=0.925\linewidth]{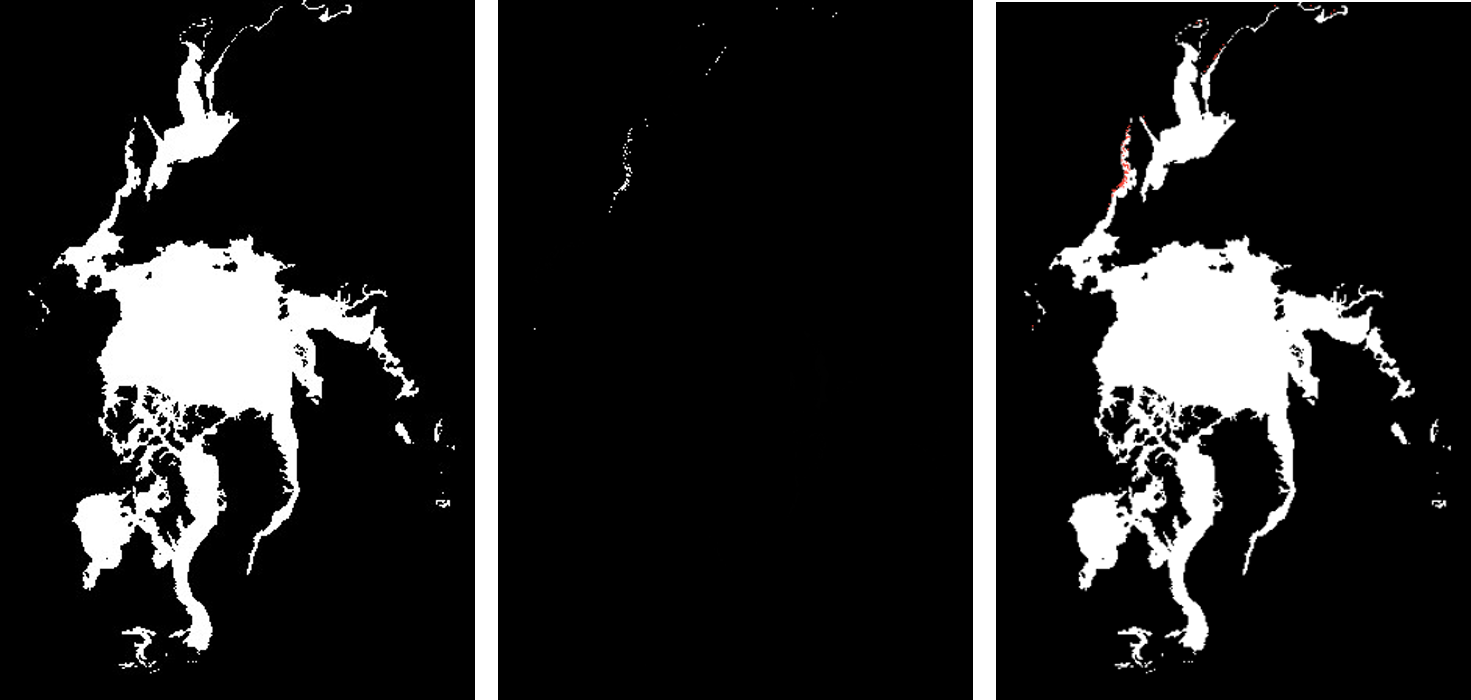}}
\caption{From left to right, where white and red represent SIE and model error (02/2016 used as example): SIE input image, 3D FFTstack residual, 3D FFTstack composite cycles}
\label{fig:SIEimage}
\end{figure}

\subsubsection{Comparative Analysis}
\label{final-comp}
The 2D GBR (de-cycled) model outperformed all the study models in all performance metrics, as shown in Table \ref{table:performance-comparison}. This model exhibited NRMSE percent decreases of 53.31\%, 37.47\%, and 40.63\% and R\textsuperscript{2} percent increases of 2.80\%, 1.12\%, and 1.43\% for the 2D GBR (original), 2D GBR (detrended), and  2D FFTstack models, respectively. Thus, the 2D GBR (de-cycled) model was overall most adept at predicting SIE.

The averaged yearly performance of each model is charted in Figure \ref{fig:SIEgraphs}B. The GBR (baseline) model exhibited a statistically significant long-term trend, suggesting its predictions are only meaningful in the short term. On the other hand, the 2D GBR (detrended), 2D FFTstack, and 2D GBR (de-cycled) models did not have statistically meaningful trends, suggesting longer-term prediction relevancy.

Figure \ref{fig:SIEgraphs}C shows a month-of-year RMSE comparison of all 2D study models. The detrended model outperformed the de-cycled model in 50\% of the months, including February, March, May, August, September, and December. Thus, the detrended model was more accurate in predicting extrema months, namely March and September. It should be noted that during these months, the detrended model exhibited only slightly better RMSE; the results were fairly comparable. In the remaining months, the de-cycled model tended to outperform the detrended model, especially in October and November. Therefore, while the 2D GBR (detrended) model may have outperformed the 2D GBR (de-cycled) model in some months by small margins, the 2D GBR (de-cycled) model markedly outperformed the detrended model in the remaining months. Though the 3D architecture was outperformed by \cite{kim2021multi}, it effectively removed most inter-yearly and seasonal cycles from the SIE images, as shown in Figure \ref{fig:SIEimage}B.

The 2D GBR (de-cycled) model outperformed all other comparable models taken from literature. In comparison to this model, \cite{ali2021sea}, \cite{ali2022benchmarking}, and \cite{kim2021multi} exhibited 50.26\%, 31.55\%, and 3.42\% higher NRMSE, respectively. Comparatively small increases in R\textsuperscript{2} were also present. In addition, when FFTstack was integrated into their frameworks, \cite{ali2021sea} and \cite{ali2022benchmarking} underwent 36.31\% and 19.72\% decreases in NRMSE, respectively.

\section{Conclusions}
This study proposes FFTstack, a process that effectively detects and/or removes temporal trends from time series data. By integrating FFTstack into a proposed 2D architecture, the 2D GBR (de-cycled) model outperformed all baseline models and existing state-of-the-art, ML, DL, and probabilistic models in RMSE, NRMSE, and R\textsuperscript{2} metrics. Due to its superior performance, the proposed model can meaningfully contribute to predictive models of related research, such as predicting sea level rise, species decline, or anomalous weather patterns. In addition, the lack of a statistically meaningful long-term linear trend suggests that this model's predictions will remain applicable in the future, unlike the predictions of several existing models. 
In future research, we will investigate changes in model performance on the daily interval and/or at longer lead times. We will also implement a DL model to the 3D architecture, which currently only supports FFTstack, in order to complete the residual framework. FFTstack also improved the performance and long-term relevancy of existing models, demonstrating its widespread compatibility and utility in similar research. In conclusion, this research represents an important step in guiding mitigation, resilience, and adaptation efforts in response to Arctic sea ice decline. For example, the 2D GBR (de-cycled) model can be used as mechanism of awareness, or a statistic guiding local and governmental action. Though more detailed and accurate data is needed, the 3D model can be used to map new shipping routes, understand changing ecosystems, and help indigenous peoples adjust to a changing environment. In short, understanding sea ice decline is the first step in understanding vastly more research, and it therefore stands as a first step to worthwhile global responses.



\section*{Acknowledgment}
Lapp acknowledges the Ingenuity Project with additional thanks to research coordinator Ms. Kowsar Ahmed and research director Dr. Nicole Rosen. Wang and Ali are supported by NSF grants OAC-1942714 and OAC-2118285.
\bibliographystyle{abbrv}
\bibliography{ieee_ref}

\end{document}